\documentclass[aps,pre,twocolumn]{revtex4-1}

\usepackage{graphicx}

\def\ddt{\partial_t}
\def\ddx{\partial_x}

\def\nablav{\mbox{\boldmath$\nabla$}}
\def\grad{\nablav}

\def\div{\grad\cdot}

\def\bea{\begin{eqnarray}}
\def\eea{\end{eqnarray}}

\newcommand{\braket}[1]{\left\langle#1\right\rangle}

\begin{document}

\title{Momentum Absorption and Magnetic Field Generation by Obliquely Incident Light}
\author{Andrea Macchi}\email{andrea.macchi@ino.cnr.it}
\affiliation{CNR, National Institute of Optics (INO), via G.~Moruzzi 1, 56124 Pisa, Italy}
\affiliation{Dipartimento di Fisica Enrico Fermi, Universit\`a di Pisa, l.go Bruno Pontecorvo 3, 56127 Pisa, Italy}
\author{Anna Grassi}\email{agrassi8@stanford.edu}
\affiliation{High Energy Density Science Division, SLAC National Accelerator Laboratory, Menlo Park, California 94025, USA}
\author{Fran\c{c}ois Amiranoff}
\affiliation{Laboratoire d'Utilisation des Lasers Intenses, Ecole Polytechnique, 91128 Palaiseau Cedex, France}
\author{Caterina Riconda}
\affiliation{LULI-UPMC: Sorbonne Universit\'{e}s, CNRS, \'{E}cole Polytechnique, CEA, 75005 Paris, France}

\date{\today}

\begin{abstract}
The partial reflection of an electromagnetic (EM) wave from a medium leads to absorption of momentum in the direction perpendicular to the surface (the standard radiation pressure) {and, for oblique incidence on a partially reflecting medium}, also in the parallel direction. This latter component drives a transverse current and a slowly growing, quasi-static magnetic field in the evanescence ``skin'' layer. Through a simple model we illustrate how EM momentum is transfered to ions and estimate the value of the magnetic field which may be of the order of the driving EM wave field, i.e. up to several hundreds of megagauss for high intensity laser-solid interactions. 
\end{abstract}

\maketitle

\section{Introduction}

When an electromagnetic (EM) wave is reflected from a medium, part of the EM momentum carried by the wave is absorbed in the medium. The flow of the momentum component perpendicular to the surface {of a medium at rest} through the surface itself corresponds to the well-known radiation pressure, and in plane geometry it is given by
\bea
P_{\perp}=(1+R)\frac{I}{c}\cos^2\theta_i \; , \qquad
\label{eq:Pperp}
\eea
where $I$ is the intensity of the wave, $\theta_i$ is the angle of incidence, {$c$ is the speed of light}, and $R$ is the reflectivity of the medium. The expression for the flow of the momentum component \emph{parallel} to the surface is maybe less familiar:
\bea
P_{\parallel}=(1-R)\frac{I}{c}\sin\theta_i\cos\theta_i \; . 
\label{eq:Ppar}
\eea
Both Eqs.(\ref{eq:Pperp}) and (\ref{eq:Ppar}) can be derived from the classical electrodynamics theory of continuum media \cite{landau-radpress}. 

Assuming a medium whose optical response is due to free electrons (e.g. a simple metal), at a ``microscopic'' level the absorption of EM momentum is due to forces on such electrons only. However, $P_{\perp}$ is promptly delivered to the background ions, since any net secular force pushing the electrons in the skin layer, i.e. in the evanescence region of the EM wave, generates a charge separation and an electrostatic field that back-holds electrons and accelerate ions. If the electrons are in mechanical equilibrium, the total electrostatic pressure exactly balances $P_{\perp}$, so that macroscopically the radiation pressure appears to be exerted on the whole medium. The pressure generated by contemporary high power laser system is the highest achievable in a laboratory and can accelerate a thin object to velocities of the order of $c$, with foreseen applications in ion accelerators \cite{esirkepovPRL04,macchiRMP13,macchiHPLSE14} {as} well as more visionary ones in interstellar propulsion \cite{marxN66,forwardJS84,meraliS16}.

The situation is different for $P_{\parallel}$, since electrons can be accelerated along the surface without generating a charge separation. As we discuss in this paper, parallel acceleration drives a surface current which in turn generates a quasistatic magnetic field and an \emph{inductive} electric field. 
Such field transfers the absorbed EM momentum to ions but, differently from the perpendicular case, cannot fully balance the nonlinear secular force on the electrons locally, so that an ambipolar electron current persists. The quasistatic magnetic field in the skin layer may reach an amplitude close to the vacuum field, which is remarkable for high intensity interactions where the value is of the order of hundreds of megagauss. 

Our aim is mainly pedagogical but the results may also be useful for the physical description of contexts when the effect is sizable, i.e for sufficiently intense laser light. 
Indeed, the onset of quasi-static transverse currents and associated magnetic fields correlated with energy absorption  
have been observed in simulations of intense laser interaction with high-density plasmas at oblique incidence\footnote{Including the cases of \emph{locally} oblique incidence due to either static or dynamic bending of the target surface.} since a long time\cite{thomsonPRL75,brunelPF88,schlegelPRE99,wilksPRL92,ruhlPRL99}. However, only rarely and fleetingly the connection to the absorption of EM momentum along the surface has been noticed explicitly \cite{kentwellPP80,brunelPF88,schlegelPRE99}. In addition, to our knowledge the self-consistent effect of the inductive field and the transfer of transverse momentum to ions have not been discussed so far. It may also be interesting to notice the similarities with the absorption of ``spin'' angular momentum by a circularly polarized laser pulse and related magnetic field generation for which the relation to dissipative absorption, the importance of induction effects and the coupling to ions were analyzied only after several years \cite{hainesPRL01,shvetsPRE02,liseykinaNJP16}.

Our present work has been stimulated by recent simulations shown in Ref.\cite{grassiPRE17} where a first summary of our results was reported. Another possible application of current interest for our analysis is the generation of transient currents on the surface of a laser-irradiated solid target as sources of terahertz pulses \cite{liAPL12,zhupPRE17}.

\section{Dynamics of momentum absorption}

\begin{figure}[b]
\begin{center}
\includegraphics[width=0.5\textwidth]{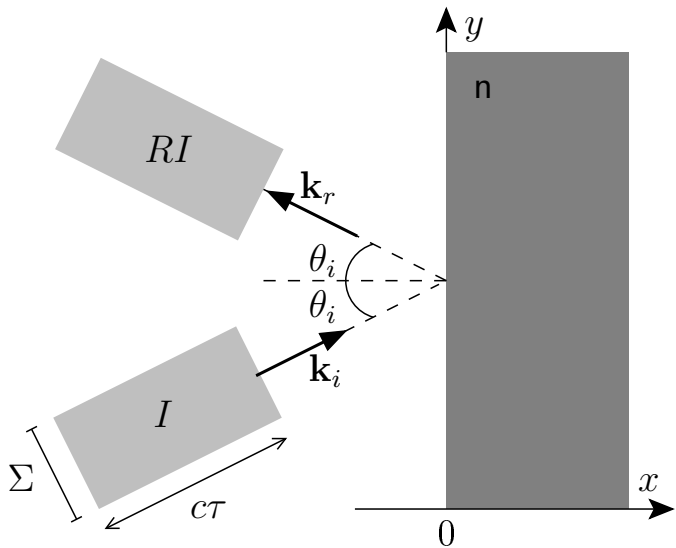}
\end{center}
\caption{Simple ``kinematical'' model: a rectangular wavepacket impinges on a medium of reflectivity $R$.\label{fig:oblref}}
\end{figure}

\subsection{Simple derivation of momentum flow}

First, for the sake of completeness and simplicity, let us derive (\ref{eq:Pperp}) and (\ref{eq:Ppar}) from simple kinematics of the reflection. 
Consider as in Fig.\ref{fig:oblref} a ``quasi-plane'', rectangularly shaped wavepacket (of finite but arbitrarily large length $c\tau$ and transverse areal section $\Sigma$) obliquely incident at an angle $\theta_i$ on an absorbing medium of infinite inertia, having reflectivity $R<1$ and a thickness much larger than the skin depth so that transmission is negligible.
From the energy-momentum conservation theorem of Maxwell's equation we know that the wavepacket delivers a flow of energy per unit surface $I \equiv |{\bf S}|$ where ${\bf S}=(c/4\pi){\bf E}\times{\bf B}$ is Poynting's vector, while the density of EM momentum is ${\bf g}={\bf S}/c^2$. Thus, the incident wavepacket of intensity $I$ contains a total momentum ${\bf p}_i=(I/c^2)(\Sigma c\tau)\hat{\bf n}_i$ where $\hat{\bf n}_i=(\cos\theta_i,\sin\theta_i)$ is the direction of incidence. Under the assumption of a quasi-monochromatic field (so that $\hat{\bf n}_i={\bf k}_i/k$ with $k=\omega/c$) the reflected wavepacket has the same shape as the incident one, and since its intensity is $RI$ the total momentum is ${\bf p}_r=R(I/c^2)(\Sigma c\tau)\hat{\bf n}_r$ with $\hat{\bf n}_r=(-\cos\theta_i,\sin\theta_i)$ according to the law of reflection. Thus, the amount of momentum $\Delta{\bf p}$ delivered to the medium is
\bea
\Delta{\bf p}={\bf p}_i-{\bf p}_r
=\frac{I}{c}\Sigma \tau\left((1+R)\cos\theta_i \, , \, (1-R)\sin\theta_i\right) \; .
\eea
The total force exerted per unit surface is obtained dividing $\Delta{\bf p}$ by the wavepacket duration $\tau$ and the area over which the wavepacket impinges, which is equal to $\Sigma/\cos\theta_i$,
\bea
{\bf P}=(P_{\perp},P_{\parallel})=\frac{\Delta{\bf p}}{c\tau (\Sigma/\cos\theta_i)} \; . 
\eea
so that (\ref{eq:Pperp}) and (\ref{eq:Ppar}) are obtained. The result is independently on $\Sigma$ and $\tau$ and thus it is appropriate to describe the limit of a plane, monochromatic wave. The flow of parallel momentum $P_{\parallel}$ is non-vanishing only in the presence of absorption ($R<1$) and for oblique incidence ($\theta_i\neq 0$).

\subsection{Ponderomotive force}

The \emph{global} pressure on the target arises from a the spatial integration of a \emph{local} secular force on the plasma electrons, i.e. a ponderomotive force (PF).
Let us relate the expression of the PF to those of the momentum flow (\ref{eq:Pperp}-\ref{eq:Ppar}). We assume all the fields and currents to have a general dependence on time such that ``fast'' and ``slow'' temporal scales can be separated, i.e.
\bea
{\bf E}({\bf r},t)=\mbox{Re}\left(\tilde{\bf E}({\bf r},t)\mbox{e}^{-i\omega t}+\mbox{c.c.}\right) \; ,
\eea
such that performing an average over the oscillation cycle (of period $2\pi/\omega$)
\bea
\braket{{\bf E}({\bf r},t)}\simeq 0 \; , \qquad \braket{\tilde{\bf E}({\bf r},t)}\simeq \tilde{\bf E}({\bf r},t) \; .
\eea
The ponderomotive force per unit volume can be thus calculated from a known distribution of oscillating EM fields as follows,
\bea
{\bf f}_p=\braket{\rho{\bf E}+\frac{\bf J}{c}\times{\bf B}}=-\div\braket{\sf T} \; ,
\label{eq:fpdivT}
\eea
where ${\sf T}={\sf T}({\bf r},t)$ is Maxwell's stress tensor
\bea
{\sf T}_{ij}=\frac{1}{4\pi}\left(\frac{{\bf E}^2+{\bf B}^2}{2}\delta_{ij}-E_iE_j-B_iB_j\right) \; .
\eea
In the case of our problem, i.e. the incidence of a plane wave on a medium extended for $x>0$ and homogeneous along $y$ and $z$, with $xy$ as the plane of incidence, we obtain
\bea
{\bf f}_p = -\ddx(\braket{\sf T}_{xx},\braket{\sf T}_{xy}) \; .
\label{eq:fpT}
\eea
Notice that because of the cycle average ${\bf f}_p$ does not depend on $y$ (all fields and currents depend on $y$ only via the phase $(k_yy-\omega t)$). 
Thus, in general ${\bf f}_p={\bf f}_p(x,t)$ where the dependence on time is slow with respect to the laser period.

Since ${\bf P}=(P_{\perp},P_{\parallel})$ corresponds to the total force per unit area on the medium, it is equal to the integral of ${\bf f}_p$ over the depth of the medium,
\bea
{\bf P}=\int_0^{\infty}{\bf f}_p\mbox{d}x=-(\braket{\sf T}_{xx},\braket{\sf T}_{xy})|_{x=0} \; .
\label{eq:P-T}
\eea
Explicit expressions for ${\sf T}$ and ${\bf f}_p$ can be obtained starting from the Fresnel formulas giving the EM field distribution in the reflection from a plane, linear medium described by a known refractive index ${\sf n}={\sf n}(\omega)$ or, equivalently, a dielectric function $\varepsilon=\varepsilon(\omega)={\sf n}^2$. This corresponds to a perturbative approach in which the response of the medium is not modified by the nonlinear forces. In this case, the fields decay exponentially inside the medium as $\exp(-x/\ell_s)$, with the evanescence length $\ell_s$ given by the generalized Snell's law, and thus ${\sf T}$ and ${\bf f}_p$ decay as $\exp(-2x/\ell_s)$.

Nevertheless, also in regimes where the field distribution is modified by nonlinear (e.g. relativistic) effects so that obtaining the field profiles is not straightforward, by keeping (\ref{eq:P-T}) as a constraint we may write in general
\bea
{\bf f}_p={\bf P}s(x) \; ,
\label{eq:fpsx}
\eea
with {$s(x)$} an evanescent function, normalized such that
\bea
\int_0^{\infty}s(x)\mbox{d}x=1 \; .
\eea

\subsection{On the energy absorption coefficient}

Differently from $P_{\perp}$, a non-vanishing $P_{\parallel}$ requires the reflectivity $R<1$, i.e. some EM energy is absorbed into the medium. 
A finite energy absorption requires the dielectric function $\varepsilon(\omega)$ to have a complex part. For a simple metal we take
\bea
\varepsilon(\omega)=1-\frac{\omega_p^2}{\omega(\omega+i\nu)} \; ,
\label{eq:dielecfunc}
\eea
with $\omega<\omega_p=(4\pi e^2 n_e/m_e)^{1/2}$ in order {for} the medium to be opaque ($n_e$ is the electron density). In Drude's model $\nu=\nu_c$, where $\nu_c$ is the friction or collision frequency related to the conductivity $\sigma=n_ee^2/m_e\nu_c$. In high intensity interactions, the collision frequency drops down because of target heating and nonlinear effects. However, sizable absorption may be due to non-collisional effects such as sheath inverse Bremmstrahlung \cite{cattoPoF77} which {still may be} included via an effective collision frequency, i.e. a suitable expression for $\nu$ in (\ref{eq:dielecfunc}). Up to this point, it is possible to solve the problem of momentum absorption and quasi-static field generation \emph{ab initio} from Maxwell's equations in a medium described by (\ref{eq:dielecfunc}), obtaining the PF from the expression of the fields and then calculating the ``slow'', quasi-static dynamics driven by the PF. 

By further increasing the laser intensity, absorption becomes dominated by the anomalous skin effect \cite{weibelPoF67,rozmusPoP96}, which is of non-local nature, and ultimately by nonlinear processes such ``vacuum heating'' (VH) \cite{brunelPRL87,brunelPF88,gibbon-book-VH} or ``${\bf J}\times{\bf B}$'' heating \cite{kruerPF85} due to electrons driven across the target-vacuum interface. Such processes may not be described via a local dielectric function (\ref{eq:dielecfunc}). However, our phenomenological model may take these processes into account via a suitable expression for the reflectivity $R$. To this aim, theories yielding some general dependence of the absorption and reflectivity coefficients as a function of laser and plasma parameters may be useful \cite{gibbonPPCF12}.

It may be worth noticing that a friction frequency $\nu$ as that appearing in (\ref{eq:dielecfunc}) may also be included in order to keep track of transient or causality effects, e.g. of the adiabatic rising of the external field, even if dissipative terms such as those due to collisions are negligible.  
Physically this describes the fact that, in a steady state as for a medium interacting with a monochromatic EM field,
the electrons of the medium have a mean oscillation energy which they acquired from the EM field as the latter was turned up; thus, the electrons must have acquired a proportional amount of momentum as well.
An example of this effect is the drift on a point charge in a plane, monochromatic EM wave when the latter is rised adiabatically \cite{macchi-book-planewave}: such drift is a ``memory'' of the momentum absorbed during the field rising, in order to reach the stade of steady oscillation in which energy and momentum are not absorbed anymore\footnote{The existence of a transverse drift after the rising of the laser pulse may be also explained by the conservation of canonical momentum.}. Coming back to our problem of interest, this implies that drifting currents and slowly varying fields may be observed also in a ``ideal'' medium (without any dissipation or energy absorption mechanism) as a memory of the rising phase; in absence of irreversible effects, such fields would vanish when the external driver (the laser pulse) is turned off. This observation may explain why secular currents are also apparent in fluid models without dissipation \cite{sudanPRL93,vshivkovPoP98}, and may be useful to the interpretation of numerical simulations of high intensity interactions.

\subsection{Steady field generation and coupling to ions}

As already mentioned, the optical properties of the medium are determined by the electron dynamics, with the ion contribution being negligible because of their large inertia. Inside the skin layer, where the EM field is evanescent, the {perpendicular} PF, i.e. the cycle-averaged component of the Lorentz force on the electrons in the direction perpendicular to the surface, has a non-vanishing positive value and pushes the electron fluid inwards. 
The pushing by the PF leads to charge separation, generating an electrostatic field which balances the PF keeping the electrons in mechanical equilibrium. Thus, ultimately the background ions feel an electric force equal to the PF, so that the momentum is transferred to the whole medium.

Along the parallel direction the situation is different. In plane geometry (i.e. neglecting boundary effects), the electrons can slide along the surface without generating any charge separation, thus the electron flow is not affected by backholding electrostatic fields. Such flow may produce a current density in the skin layer and, in turn, a magnetic field. However, the rise of the magnetic field generates an \emph{inductive} electric field, which decelerates electrons and accelerates ions. This is the basic mechanism by which also the {parallel} momentum may be eventually transfered to the ions.

The fact that, contrary to what happens for the {perpendicular} component, the sum of {the parallel components of} the PF and of the electric force does not vanish locally is due to different screening lengths in the evanescence region. For a simple metal described by (\ref{eq:dielecfunc}) in a quasi-linear regime, the slowly varying electric field is screened on a distance $\sim c/\omega_p$ (the collisionless skin depth), while since the PF is proportional to the square of the fields its screening length {$\sim (c/2\omega_p)$}. This leads to the generation of ambipolar electric field and current density in the evanescence region, where the magnetic field is localized.

\section{Fluid modeling}

We now describe the generation of quasi-static electric and magnetic fields following the absorption of {parallel} momentum using a ``minimal'' model based on a classical fluid description for the electrons. Assuming the PF to be determined by the optical properties of the medium, we consider the ``slow'' electron dynamics driven by the PF. In general, the equation for the slow component of the electron velocity (assumed to be non-relativistic) is 
\bea
\frac{\mbox{d}{\bf u}_s}{\mbox{d}t}&=& 
-\frac{e}{m_e}\left({\bf E}_s+\frac{{\bf u}_s}{c}\times{\bf B}_s\right) \nonumber \\ 
& &+\frac{1}{n_em_e}{\bf f}_p-\nu_s {\bf u}_s
-\frac{\grad{\sf P}}{n_e} 
\; ,
\label{eq:sloweq}
\eea
where $\nu_s$ is a friction coefficient and ${\sf P}$ is a pressure term. Eq.(\ref{eq:sloweq}) is coupled to Maxwell's equations by the current density ${\bf J}_s=-en_e{\bf u}_s$. 

We assume the medium to be inhomogeneous along $x$ and the PF to have $x$ and $y$ components ${\bf f}_p=(f_{px},f_{py})$. As described above, along $x$ the PF is locally balanced by the electrostatic field, and $u_{sx}=0$. From now on we restrict our analysis on the motion along $y$. The relevant field and current components are $E_{sy}$, $B_{sz}$, and $J_{sy}$. As stated above, $E_{sy}$ is of inductive nature. In addition, we assume the displacement current to be neglible. Thus, we have the equations
\bea
\frac{\mbox{d}{u_{sy}}}{\mbox{d}t}&=&-\frac{e}{m_e}E_{sy}+\frac{1}{n_em_e}f_{py}-\nu_su_{sy} \; , \label{eq:euls} \\
\ddx E_{sy}&=&-\frac{1}{c}\ddt B_{sz} \; , \label{eq:fars} \\
\ddx B_{sz}&=&-\frac{4\pi}{c}J_{sy}=\frac{4\pi e n_e}{c}u_{sy} \label{eq:amps}\; .
\eea

\subsection{Collisionless regime}

The simplest but yet instructive situation in which we solve Eqs.(\ref{eq:euls}-\ref{eq:fars}-\ref{eq:amps}) is the one in which we neglect the friction term ($\nu_s=0$) and we consider the electron density as homogeneous in the $x>0$ region, i.e. $n_e=n_0\Theta(x)$. Besides obvious reasons of simplicity, similar conditions may be created by laser-solid interactions using femtosecond, high contrast pulses \cite{gibbonNP07} at intensities typically exceeding $10^{16}~\mbox{W cm}^{-2}$, since the solid material is isochorically heated to high temperatures and the collision frequency drops down. Below intensities of some $10^{18}~\mbox{W cm}^{-2}$, nonlinear effects such as ``relativistic'' transparency and profile steepening by radiation pressure play a modest role and the optical properties may be quite well described by the dielectric function (\ref{eq:dielecfunc}). 

The final, and possibly the crudest approximation we make is to linearize the total time derivative in  Eq.(\ref{eq:euls}), i.e. to assume $\ddt u_s \gg u_s\ddx u_s$. As it will appear \emph{a posteriori}, this requires $(\omega_p\tau)(u_s/c)\ll 1$, where $\tau$ is the typical growth time of the quasi-static field which will turn to be the laser pulse duration.

Within the above assumptions, it is straightforward to obtain the following equation for $E_{sy}=E_{sy}(x,t)$:
\bea
\left(\ddx^2-\frac{\omega_p^2}{c^2}\right)E_{sy}=-\frac{4\pi e}{m_ec^2}f_{py} \; .
\label{eq:helmE}
\eea
If we take, consistently with a perturbative approach and Eq.(\ref{eq:dielecfunc}), $f_{py}=f_{py}(0,t)\mbox{e}^{-2x/\ell_s}$ where {$\ell_s=c(\omega_p^2-\omega^2\cos^2\theta)^{-1/2}$}, the solution of (\ref{eq:helmE}) is
\bea
E_{sy}&=&-\frac{4\pi e}{m_ec^2}\frac{\ell_s^2}{4-\omega_p^2\ell_s^2/c^2}f_{py}(0,t)\mbox{e}^{-2x/\ell_s} \nonumber \\
 & & +E_{\rm h}(t)\mbox{e}^{-\omega_p x/c} \; ,
\label{eq:Esy0}
\eea
with the function $E_{\rm h}(t)$ to be determined by boundary conditions. 
Heuristically, the appearance of two different spatial scales is due to the fact that the ``forcing'' length ($\ell_s/2$) is different from the ``screening'' length ($c/\omega_p$). 

For further simplification we assume $\omega_p\gg\omega$,  as it is the case for optical laser frequencies and solid-density targets, so that (\ref{eq:Esy0}) reduces to
{
\bea
E_{sy}&=&-\frac{1}{3en_0}f_{py}(0,t)\mbox{e}^{-2\omega_px/c}+E_{\rm h}(t)\mbox{e}^{-\omega_p x/c} \; .
\label{eq:Esy1}
\eea
}
{
Using Eq.(\ref{eq:fars}) we obtain from (\ref{eq:Esy1}) the magnetic field
\bea
\ddt B_{sz} &=&-\frac{2\omega_p}{3en_0}f_{py}(0,t)\mbox{e}^{-2\omega_px/c} \nonumber \\ & &+\omega_p E_{\rm h}(t)\mbox{e}^{-\omega_p x/c} \; .
\eea
Within the quasi-static approximation underlying Eq.(\ref{eq:amps}), it may seem reasonable to infer that the magnetic field will be generated only in the $x>0$ region where the current flows, i.e. to assume $B_{sz}(x=0,t)=0$. Such boundary condition, which will be further discussed below, yields $E_{\rm h}(t)=2f_{py}(0,t)/3en_0$, so that 
\bea
B_{sz} &=&\frac{2\omega_p}{3en_0}
\left(\mbox{e}^{-\omega_p x/c}-\mbox{e}^{-2\omega_px/c}\right) \nonumber \\ 
& & \times \int_0^tf_{py}(0,t')\mbox{d}t' \; .
\label{eq:Bsz0}
\eea
}
The electric field corresponding to Eq.(\ref{eq:Bsz0}) is 
\bea
E_{sy} =\frac{1}{3en_0}
\left(2\mbox{e}^{-\omega_p x/c}-\mbox{e}^{-2\omega_px/c}\right)f_{py}(0,t) \; .
\label{eq:Esy0.1}
\eea
Notice that $E_{sy}>0$ as expected, since it should accelerate electrons along the $-\hat{\bf y}$ direction, opposite to $f_{py}$.  
{If $f_{py}(0,t)$ is a smooth bell-shaped function (e.g. a Gaussian opulse)of duration $\tau$, the ratio between the amplitudes of $E_{sy}$ and $B_{sz}$ is $\sim (\omega_p\tau)^{-1}\ll 1$. However, albeit small $E_{sy}$ does not vanish at $x=0$, which shows that assuming zero quasistatic fields on the vacuum side is an approximation (see below).}

\begin{figure}
\begin{center}
\includegraphics[width=0.5\textwidth]{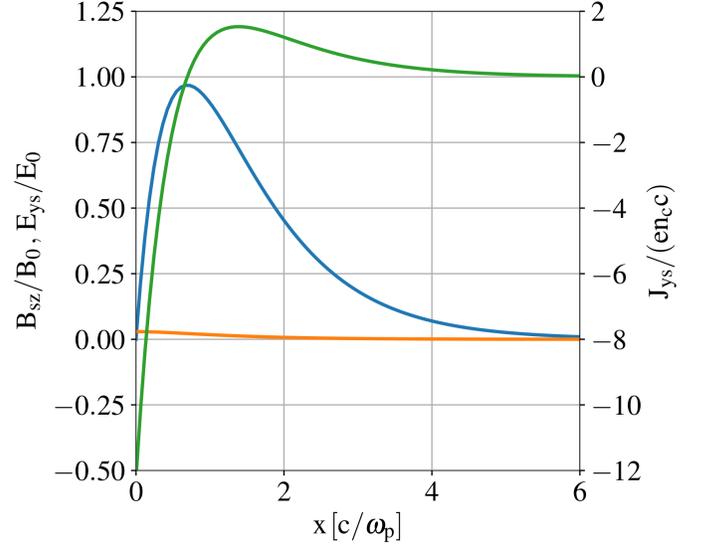}
\caption{The spatial profiles of $B_{sz}$ (blue), $E_{ys}$ (orange) and $J_{ys}$ (green). The values of the fields are normalized to the laser amplitude in vacuum $E_0=B_0$ and are plotted at the end of a laser pulse with Gaussian profile and duration $\tau_L=10T_L$. Other parameters are $a_0=1$, $n_e/n_c=10$, and $R=0.25$. \label{fig:fieldprofiles}}
\end{center}
\end{figure}

The peak magnetic field is at $x=(c/\omega_p)\ln 2$, for which $(\mbox{e}^{-\omega_px/c}-\mbox{e}^{-2\omega_p x/c})=1/4$. Now we have, using Eqs.(\ref{eq:Pperp}-\ref{eq:Ppar}) and (\ref{eq:fpT})
\bea
f_{py}(0,t)=\frac{2\omega_p}{c^2}I(t)\frac{1-R}{2}\sin(2\theta_i) \; .
\eea
The fluence of the laser pulse may be estimated as
\bea
\int_0^{\infty}I(t')dt' \simeq I_L\tau_L \; ,
\eea
with $I_L$ the average intensity and $\tau_L$ the duration. 
We also write $I_L=m_en_cc^3a_0^2$ with $n_c$ the cut-off density, such that $n_0/n_c=\omega_p^2/\omega^2$, and $a_0$ the dimensionless laser amplitude such that relativistic effects are small for $a_0<1$. Introducing $B_0=m_ec\omega/e$ and the laser period $T_L=2\pi/\omega$, and rearranging the above expressions a bit we get for the maximum $B$-field
{
\bea
\mbox{max}(B_{sz}) \simeq \frac{4\pi}{3}B_0a_0^2\frac{\tau_L}{T_L}(1-R)\sin(2\theta_i) \; .
\label{eq:Bmax}
\eea 
}
Notice that $B_0a_0=B_L$, the amplitude of the laser field. With parameters such as $a_0 \sim 1$, $\tau_L/T_L \sim 10$ (i.e. a $\sim$30~fs, $10^{18}~\mbox{W cm}^{-2}$ pulse for $\lambda=0.8\mu\mbox{m}$), and {$R=0.25$},
we already obtain quasi-static fields of the order of the laser magnetic field value (see Fig.\ref{fig:fieldprofiles}).

{
Actually, $B_{sz}$ does not necessarily vanish on the vacuum side. A more rigorous approach is to determine $E_{\rm h}(t)$ by the ``radiative'' boundary condition, i.e. by matching the quasi-static EM fields with those of an EM wave propagating in vacuum and along the $(-x)$ direction, for which $E_{sy}(x,t)=-B_{sz}(x,t)$. This yields the boundary condition $E_{sy}(0,t)=-B_{sz}(0,t)$ from which we obtain an equation for $E_{\rm h}(t)$,
\bea
(\ddt+\omega_p)E_{\rm h}(t)=\frac{1}{3en_0}\left(\ddt f_{py}(0,t)+2\omega_pf_{py}(0,t)\right) \; ,
\label{eq:Eom}
\eea
with general solution
\bea
E_{\rm h}(t)&=&E_{\rm h}(0)\mbox{e}^{-\omega_p t}+\frac{1}{3en_0}\int_0^t\mbox{e}^{-\omega_p (t-t')} \nonumber \\ 
& &\times 
\left(\ddt f_{py}(0,t')+2\omega_pf_{py}(0,t')\right)\mbox{d}t' \; .
\label{eq:EomSOL}
\eea
Assuming that the quasistatic fields rise for $t>0$, we may pose $E_{\rm h}(0)=0$. In addition, the temporal variation of $f_{py}$ is on the scale of the laser pulse duration, i.e. several times the period $2\pi/\omega$, thus $\ddt f_{py}<\omega f_{py}\ll \omega_pf_{py}$. Thus we may neglect the first term in the integrand and assume the second one to be constant so that it can taken out of the integral. In this way we obtain $E_{\rm h}(t)=2f_{py}(0,t)/3en_0$ again, proving that $B_{sz}(x=0,t)=0$ is accurate enough as a boundary condition.} 
{This is confirmed by evaluating (\ref{eq:EomSOL}) for a Gaussian temporal profile, which shows that $B_{sz}(x=0,t)\ll \mbox{max}(B_{sz})$. The ambipolar profile of the current density $J_{sy}=(c/4\pi)\ddx B_{sz}$ is also shown.}

{
The above discussion of radiative boundary conditions indicates that a pulse of radiation with duration $\sim \tau_L$ will be emitted back from the plasma because of the transient nature of the current. Although the effect is negligible for what concerns momentum absorption, it is of possible interest for diagnostic purposes and generation of long wavelength radiation. 
From Eq.(\ref{eq:EomSOL}) and (\ref{eq:Bmax}) we roughly estimate that the amplitude of the ``long'' pulse may be up to $\sim B_L/(\omega_p\tau_L)$, i.e. some $10^{-3}$ times the field of the driving laser for solid densities ($\omega_p \sim 10\omega$). Actually, one would probably observe a longer emission, determined by the decay of the quasi-static current $J_{sy}$ generated during the interaction.
}

For a PF with a generic spatial profile $s(x)$ as in Eq.(\ref{eq:fpsx}), the solution of Eq.(\ref{eq:helmE}) can be written as
\bea
E_{sy}=-\frac{4\pi e}{m_ec^2}\int_{-\infty}^{+\infty}\mbox{d}x'
G_H(x-x') P_{\perp}(t)\bar{s}(x') \; ,
\eea
where the Green's function
\bea
G_H(x)=\frac{c}{2\omega_p}\mbox{e}^{-\omega_p|x|/c}
\eea
and
\bea
\bar{s}(x)=\Theta(x)s(x)+\Theta(-x)s(-x) \; .
\eea
However, this solution is of limited usefulness because deviations of $s(x)$ from a simple exponential profile would be due to nonlinear effects appearing at higher intensities, such as density profile modification and relativistic corrections, which have been already neglected on the route to Eq.(\ref{eq:helmE}). For very high magnetic fields, saturation effects such as those due to finite Larmor radius could also be important. 

Ultimately, one should also consider that our simple model is based on fluid equations, with any kinetic phenomenon ``hidden'' into the expression for the reflectivity. Consistently with this approach, it is assumed that the EM momentum along the surface is transfered via the secular PF to the bulk of electrons, which possibly results in a large current density but with small drift velocity: the corresponding magnetic field may have a guiding effect for ``fast'' electrons in the high-energy tail of the distribution function. Notice that previous investigations of surface magnetic fields in intense laser-solid interaction have attributed their generation to the ``fast'' electron current, although the mechanism for electron acceleration to high energies in the direction tangent to the surface is not clear  {(at least in the absence of surface waves, see e.g. Ref.\cite{fedeliPRL16})}.

Despite all the above mentioned limitations, we argue that Eq.(\ref{eq:Bmax}) may still provide some useful estimates at intensities high enough to violate some underlying assumptions (as it was done in Ref.\cite{grassiPRE17}) since the model is mostly based on conservation laws. More importantly, the derivation of Eq.(\ref{eq:Bmax}) highlights the important role of the inductive field and of different spatial scales between the PF and quasi-static fields.

\subsection{Ohmic conductor}

As an example of a different regime we consider a Ohmic conductor for which $\nu\gg\omega$ in Eq.(\ref{eq:dielecfunc}) and $\nu_s=\nu$ in Eq.(\ref{eq:euls}). Posing ${\mbox{d}{u_{sy}}}/{\mbox{d}t}=0$ in Eq.(\ref{eq:euls}) we obtain an inhomogeneous diffusion equation for the quasi-static magnetic field,
\bea
\ddt B_{sz}=D\ddx^2B_{sz} {-} \frac{c}{en_0}\ddx f_{py} \; ,
\label{eq:diffB}
\eea
where $D=\nu c^2/\omega_p^2$. Thus, the magnetic field generated in the skin layer diffuses into the deeper layers of the medium. 
In this regime, $f_p(x,t) \simeq f_p(0,t)\mbox{e}^{-2x/\ell_c}\cos^2(x/\ell_c)$ where {$\ell_c=(c/\omega_p)(2\nu/\omega)^{1/2}$} is the resistive (or collisional) skin depth. 

The solution of (\ref{eq:diffB}) can be written with the help of the Green's function
\bea
G_D(x,t)=\frac{\Theta(t)}{(4\pi Dt)^{1/2}}\exp\left(-\frac{x^2}{4Dt}\right) \; .
\eea
{
In order to take the boundary condition at $x=0$ into account, we extend the target up to $x=-\infty$ and ``prolongate'' the source term 
${\cal S}(x,t)=-({c}/{en_0})(\ddx f_{py})(x,t)$
antisymmetrycally on the $x<0$ side. We thus obtain 
\bea
B_{sz}(x,t)
&=& \int_{0}^t\mbox{d}t'\int_{-\infty}^{+\infty}\mbox{d}x' G_D(x-x',t-t') 
\nonumber \\
& & \times 
\left[ \Theta(x'){\cal S}(x',t')-\Theta(-x')S(-x',t')\right] \; .
\label{eq:B-G}
\eea

The Green's function becomes wider in space than the source term when  
$4Dt>\ell_c^2$, i.e. for times $t>\ell_c^2/4D=1/2\omega$, which is shorter 
than one laser cycle. This implies that an any relevant time the magnetic field growth will be limited by magnetic diffusion, and that the source term in (\ref{eq:B-G}) will be localized with respect to the Green's function $G_D$. This allows us to perform a rough, order-of-magnitude estimate of (\ref{eq:B-G}) by taking only the leading source term in (\ref{eq:B-G}) into account:
\bea
B_{sz}(x,t)
&\sim & \int_{0}^t\mbox{d}t'\int_{-\infty}^{+\infty}\mbox{d}x'G_D(x-x',t-t') 
\nonumber \\
& & \times 
\frac{c}{en_0}f_{py}(0,t')\frac{\mbox{d}}{\mbox{d}x'}\mbox{e}^{-2|x'|/\ell_s}
\nonumber \\
&\simeq& -\frac{c}{en_0}\int_{0}^t\mbox{d}t'f_{py}(0,t')
\nonumber \\
& & \times \int_{-\infty}^{+\infty}\mbox{d}x'\mbox{e}^{-2|x'|/\ell_s}\frac{\mbox{d}}{\mbox{d}x'}G_D(x-x',t-t') 
\nonumber \\
&\simeq&
\frac{2c\ell_sx}{\sqrt{\pi}en_0}\int_{0}^t\mbox{d}t'\frac{f_{py}(0,t')\mbox{e}^{-{x^2}/{4D(t-t')}}}{(4 D(t-t'))^{3/2}}
 \; .
\label{eq:B-G2}
\eea
For a laser pulse of sufficiently short duration $\tau_L$ and peaked in time, the maximum field will be reached near the pulse peak, i.e. 
\bea
B_{sz}(x,\tau_L) \sim \frac{2c\ell_sx}{\sqrt{\pi}en_0}\frac{\mbox{e}^{-{x^2}/{4D\tau_L}}}{(4D\tau_L)^{3/2}}
\int_{0}^{\tau_L}\mbox{d}t'{f_{py}(0,t')} \; .
\eea
With respect to Eq.(\ref{eq:Bsz0}), the reduction factor in $B_{sz}$ is of the order $\sim (\omega/\nu)^{1/2}(T_L/\tau_L)$ which, for the collision-dominated Ohmic regime, is typically $\sim 10^{-2}$ or smaller. Considering also the scaling with $a_0$, we conclude that the static field is very small with respect to the laser field $B_L$ in this regime.

}

\section{Conclusions}

We analyzed the basic mechanism by which the absorption of parallel (tangential to the surface) momentum in laser-solid interactions at oblique incidence drives the generation of a quasi-static magnetic field in th skin layer and can be eventually delivered to ions. The essentials of the mechanism are the inductive nature of the slowly varying electric field which opposes the transverse ponderomotive force and the difference between the screening and the forcing lengths. By analyzing two limiting cases, chosen for analytical feasibility, we infer that while in an Ohmic conductor the quasi-static magnetic field is strongly quenched by resistive and diffusion effects, in a collisionless plasma the magnetic field can is confined in the skin layer and reach values of the order to the laser field. While a quantitative analysis of realistic regimes should include several effects neglected in our example models and would likely require numerical simulations, we argue that the qualitative scenario remains essentially the same as outlined in the present paper.

\bibliography{paper_curry}

\acknowledgments
A.G. and C.R. acknowledge useful discussions with M.~Grech.

\end{document}